\documentclass[notitlepage,onecolumn,showpacs]{revtex4-1}
\usepackage[utf8]{inputenc}
\usepackage{amsmath}
\usepackage{amsfonts}
\usepackage{amssymb}
\usepackage{times,fullpage}
\usepackage{comment}
\usepackage{array}

\begin{document}

\newcommand{\nwc}{\newcommand}
\nwc{\vs}{\vspace}
\nwc{\hs}{\hspace}
\nwc{\la}{\langle}
\nwc{\ra}{\rangle}
\nwc{\nn}{\nonumber}
\nwc{\Ra}{\Rightarrow}
\nwc{\wt}{\widetilde}
\nwc{\lw}{\linewidth}
\nwc{\ft}{\frametitle}
\nwc{\ben}{\begin{enumerate}}
\nwc{\een}{\end{enumerate}}
\nwc{\bit}{\begin{itemize}}
\nwc{\eit}{\end{itemize}}
\nwc{\dg}{\dagger}
\nwc{\mA}{\mathcal A}
\nwc{\mD}{\mathcal D}
\nwc{\mB}{\mathcal B}

\nwc{\Tr}[1]{\underset{#1}{\mbox{Tr}}~}
\nwc{\pd}[2]{\frac{\partial #1}{\partial #2}}
\nwc{\ppd}[2]{\frac{\partial^2 #1}{\partial #2^2}}
\nwc{\fd}[2]{\frac{\delta #1}{\delta #2}}
\nwc{\pr}[2]{K(i_{#1},\alpha_{#1}|i_{#2},\alpha_{#2})}
\nwc{\av}[1]{\left< #1\right>}

\nwc{\zprl}[3]{Phys. Rev. Lett. ~{\bf #1},~#2~(#3)}
\nwc{\zpre}[3]{Phys. Rev. E ~{\bf #1},~#2~(#3)}
\nwc{\zpra}[3]{Phys. Rev. A ~{\bf #1},~#2~(#3)}
\nwc{\zjsm}[3]{J. Stat. Mech. ~{\bf #1},~#2~(#3)}
\nwc{\zepjb}[3]{Eur. Phys. J. B ~{\bf #1},~#2~(#3)}
\nwc{\zrmp}[3]{Rev. Mod. Phys. ~{\bf #1},~#2~(#3)}
\nwc{\zepl}[3]{Europhys. Lett. ~{\bf #1},~#2~(#3)}
\nwc{\zjsp}[3]{J. Stat. Phys. ~{\bf #1},~#2~(#3)}
\nwc{\zptps}[3]{Prog. Theor. Phys. Suppl. ~{\bf #1},~#2~(#3)}
\nwc{\zpt}[3]{Physics Today ~{\bf #1},~#2~(#3)}
\nwc{\zap}[3]{Adv. Phys. ~{\bf #1},~#2~(#3)}
\nwc{\zjpcm}[3]{J. Phys. Condens. Matter ~{\bf #1},~#2~(#3)}
\nwc{\zjpa}[3]{J. Phys. A ~{\bf #1},~#2~(#3)}
\nwc{\zpjp}[3]{Pramana J. Phys. ~{\bf #1},~#2~(#3)}

\title{Derivation of the not-so-common fluctuation theorems}
%\author{Sourabh Lahiri and A. M. Jayannavar}
%\date{}
\author{Sourabh Lahiri$^1$} 
\email{lahiri@kias.re.kr}
 \author{A. M. Jayannavar$^2$}
\email{jayan@iopb.res.in}
\affiliation{$^1$Korea Institute for Advanced Study, 85 Hoegiro, Dongdaemun-gu, Seoul 130-722, Republic of Korea\\
$^2$Institute of Physics, Sachivalaya Marg, Bhubaneswar 751005, India}

\begin{abstract}
The detailed fluctuation theorems of the exact form $P(A)/P(-A)=e^A$ exist only for a handful of variables $A$, namely for work (Crooks theorem), for total entropy change (Seifert's theorem), etc. However, the so-called modified detailed fluctuation theorems can be formulated for several other thermodynamic variables as well. The difference is that the modified relations contain an extra factor, which is dependent on $A$. This factor is usually an average of a quantity $e^{-B}$, where $B\neq A$, with repect to the conditional probability distribution $P(B|A)$. 
%It is in general feasible, although difficult, to verify these relations experimentally, owing to the difficulty in constructing the conditional probability distribution $P(B|A)$. 
The corresponding modified integral fluctuation theorems also differ from their original counterparts, by not having the usual form $\av{e^{-A}}=1$. The generalization of these relations in presence of feedback has been discussed briefly. The results derived here serve to complement the already existing results in fluctuation theorems. The steps leading to the quantum version of these derivations have been outlined in the appendix.
\end{abstract}

\maketitle
\section{Introduction}

The recently discovered fluctuation theorems (FTs) consist of a group of relations that hold for a nonequilibrium system, no matter how far the system has been driven away from equilbrium  \cite{sei05,sei08,jar97,jar97a,eva02,har07,kur07,zon03_prl,
zon04,dha04,cro99,
kur98,cro98}. They come in two different forms. The first form is the detailed fluctuation theorem (DFT) that relates the probability distributions of thermodynamic variable $A$ observed in the forward and time-reversed processes (see below), respectively. 
The generic form  for a DFT is $P_f(A)/P_r(-A)=e^A$. The subscripts $f$ and $r$ denote forward (with external drive $\lambda(t)$) or reverse (with external drive $\lambda(\tau-t)$) processes, if the process is carried out from time $t=0$ to $t=\tau$. The thermodynamic variables  that are usually involved consist of heat, work or entropy. 
The second form is known as the integral fluctuation theorem (IFT), which can often be obtained from the corresponding DFT by integrating over all values of the involved thermodynamic variable. The IFT is given by the generic form $\av{e^{-A}}_f=1$, when the $\av{\cdots}_f$ denote ensemble averaging over the phase space trajectories generated in the forward process.
The second law can be derived as a corollary from these theorems. There have been several excellent reviews on FT in recent years \cite{rit03,jar10_arcmp,han11,sei12}.

Exact fluctuation theorems have been derived for the work  $W$ done on the system \cite{jar97,jar97a,cro98,sei12}, and for  the total entropy change $\Delta s_{tot}$ in the system and the surrounding medium \cite{sei05,sei08,cro99,sei12}. 
Several important thermodynamic variables like the system entropy change $\Delta s$, internal energy change $\Delta E$ and dissipated heat $Q$, do not have exact fluctuation theorems (in some cases, $Q$ does follow a DFT only when the time of observation  is very large \cite{zon03_prl}).
In this article, we follow the technique used in \cite{han09_jsm,gar12} to  show that even these variables do follow a modified form of fluctuation theorems, like the modified DFT (MDFT) and  modified IFT (MIFT).  The difference of these modified relations from the usual DFT and IFT has been discussed later. The derived results complement the  known fluctuation relations in literature.  Apart from deriving the modified FTs for $\Delta s$,  $\Delta E$ and  $Q$,  we have later introduced a new variable $\Delta D$ that also satisfies an MDFT and an MIFT.   Such modifed relations are quite uncommon in the  literature, perhaps due to the fact that they do not lead to a useful inequality like the second law.

We essentially use a single relation for all the derivations, namely \cite{sei05,sei08,sei12}
\begin{align}
\frac{P_f[X]}{P_r[\bar X]} = e^{\Delta s_{tot}[X]},
\label{eq:DFT_traj0}
\end{align}
where $X$ and $\bar X$ are the forward and time-reversed trajectories in phase space, respectively. 
The subscripts $f$ and $r$, as mentioned earlier, imply the time-dependence of the external drive, the forward process being characterized by the drive $\lambda(t)$ and the reverse process being characterized by $\lambda(\tau-t)$, respectively, where $\tau$ is the total time of observation. Thus, $P_f[X]$ gives the probability of observing the forward trajectory $X$  in the forward process, while $P_r[\bar X]$ is the probability for obtaining the corresponding reverse trajectory in the reverse process.

To prove the MIFTs, the above equation is sufficient. However, to derive the MDFTs, one needs to convert the ratio of trajectories appearing in the LHS of eq. \eqref{eq:DFT_traj0}, into the ratios of joint probabilities of variables whose sum equals $\Delta s_{tot}$ (see below). To do this,
 we simply use the definition for a joint probability. For example, suppose two variables $A$ and $B$ exist such that we have $A[X]+B[X]=\Delta s_{tot}[X]$. $A[X]$ and $B[X]$ can in general be path variables. Then one can define the joint probability of observing $A[X]=\mathcal{A}$ and $B[X]=\mathcal{B}$, in the forward process, as
\begin{align}
P_f(\mathcal{A},\mathcal{B}) = \av{\delta(\mathcal{A}-A[X])~\delta(\mathcal{B}- B[X])},
\end{align}
where the averaging has been carried out over all paths.   Similar definitions can be used for the joint probabilities for time-reversed variables (denoted by overhead bars):
\begin{align}
P_r(\bar{\mathcal{A}},\bar{\mathcal{B}}) = \av{\delta(\bar{\mathcal{A}}-\bar A[X])~\delta(\bar{\mathcal{B}}-\bar B[X])}.
\end{align}
We will usually deal with variables that are related to their time-reversed counterparts through the relations $\bar{\mathcal{A}} = \epsilon_A\mathcal{A}$, and $\bar{\mathcal{B}} = \epsilon_B\mathcal{B}$. Here, the time-parity operator $\epsilon_A$ and $\epsilon_B$ can take up the values $\pm 1$, depending on whether the observable is even  or odd  under time reversal \cite{gar12}. In this work, we will deal entirely with variables that have odd parity with respect to time-reversal, under suitably defined conditions for the forward and reverse processes.
For instance,  quantities like heat, work or internal energy change always change sign under time-reversal. However, some quantities like system entropy change only do so if the process begins and ends with the system being in nonequilibrium steady states (NESS), or in equilibrium states \cite{cro99}.

\section{The basic starting equation, and some notations}

We now check how to convert the equation \eqref{eq:DFT_traj0} into a relation of the form:
\begin{align}
\frac{P_f(\mA,\mB)}{P(\epsilon_A\mA,\epsilon_B\mB)} = e^{\Delta s_{tot}},
\label{eq:main}
\end{align}
provided the total entropy can be written as $\Delta s_{tot}=\mA+\mB$. We have,
\begin{align}
P_f(\mathcal{A},\mathcal{B}) &= \av{\delta(\mathcal{A}-A[X])~\delta(\mathcal{B}- B[X])} \nn\\
&= \int \mathcal{DX}~ P_f[X]\delta(\mathcal{A}-A[X])~\delta(\mathcal{B}- B[X]) \nn\\
&= \int \mathcal{DX}~ P_r[\bar X]e^{A[X]+B[X]}~\delta(\mathcal{A}-A[X])~\delta(\mathcal{B}- B[X]) \nn\\
&= e^{\mA+\mB}\int \mathcal{DX}~ P_r[\bar X]~\delta(\epsilon_A\bar{\mA}-A[\bar X])~\delta(\epsilon_B\bar{\mB}- B[\bar X]) \nn\\
&= e^{\Delta s_{tot}}~ P_r(\epsilon_A\bar{\mA},\epsilon_B\bar{\mB}),
\label{eq:basic}
\end{align}
which is same as eq. \eqref{eq:main}. Although the above derivation has been provided for two variables $\mathcal{A}$ and $\mathcal{B}$, the result can be extended to any number of variables, the summation of which gives $\Delta s_{tot}$ \cite{gar12}. The derivations used will essentially be based on this result, which shows how the ratio between forward and reverse trajectories be converted to the ratio between the distributions for the variables obtained in the two processes. 

In our derivations, we will consider three cases: (i) the specific case when the system begins at thermal equilibrium in either process,  (ii) the general case when the initial distribution is arbitrary, and (iii) the special case when the system begins and ends in  steady states. In case (i), we note that $\Delta s_{tot}=\beta(W-\Delta F)$, which is the dissipated work  during the process, $\Delta F$ being the change in the free energy. The following notations have been used extensively:
\begin{enumerate}

\item The symbol $A[X]$ indicates the path variable $A$, while the symbol $A$ (without the path dependence) implies the specific value $A[X]=A$.

\item In general, the symbol $\av{A}$ can be interpreted in two ways: (i) it is the average of $A[X]$ over all trajectories, or (ii) it is the average of $A$ with respect to the distribution $P(A)$ generated in the process.

\item $\av{A|B}$ implies that $A$ has been averaged with respect to the conditional probability $P(A|B)$.

\item $\av{A}'_f$ ($\av{A}'_r$) would imply that the average of $A$ has been computed for the forward (reverse) process, when the system is initially at thermal equilibrium.

\item $\av{A}^{ss}_f$ ($\av{A}^{ss}_r$) would signify that the average of $A$ has been computed for the forward (reverse) process, when the system begins and ends in steady states (or in equilibrium states) during the process.

\item Simply writing $\av{A}_f$ ($\av{A}_r$) without the prime symbol, means that the initial distribution for the forward (reverse) process can be arbitrary.

\item The probability distribution $P'_f(A)$ gives probability distribution of $A$, for the forward process that starts from equilibrium. $P_f^{ss}$ gives the same when the at initial and final times the system is in NESS (or at equilibrium), while simply writing $P_f(A)$ does not impose any restriction on the initial or the final state distributions. Similar definitions hold for the probabilities computed for the reverse process.
\end{enumerate}

\section{Derivations of MDFTs for $Q$, $\Delta s$ and $\Delta E$}

\subsection{MDFT for $Q$}

We consider a mesoscopic system in contact with a heat bath at inverse temperature $\beta=1/T$. The initial distribution of the system state is $p_0(x_0)$. It is now subjected to an external perturbation $\lambda(t)$, which drives the system out of equilibrium. The process is carried out from time $t=0$ to time $t=\tau$. At $t=\tau$ the system states follow the distribution $p_1(x_\tau)$.
All the results provided in this section essentially follow from the DFT for total entropy at the trajectory level:
\begin{align}
\frac{P_f[X]}{P_r[\bar X]} =e^{\Delta s_{tot}[X]} =  e^{\beta Q[X]+\Delta s(x_0,x_\tau)}.
\label{eq:DFT_traj}
\end{align}
Here, we have divided the total entropy change into two parts: one is the entropy change of the medium (given by $\beta Q$), and the other part is the entropy change of the system (given by $\Delta s = \ln [p(x_0)/p_(x_\tau)]$) \cite{sei05,sei08}. 
We note that the thermodynamic quantities $Q$ (dissipated heat), $\Delta E$ (internal energy change) and $W$ (work done) all switch signs under time-reversal: $\epsilon_Q=\epsilon_{\Delta E}=\epsilon_W=-1$. 
In contrast, the system entropy change $\Delta s$ has $\epsilon_{\Delta s}=-1$ \emph{only when the system begins and ends in a steady state} \cite{sei05,sei08,cro99}.

For a system beginning from an equilibrium state in either (forward and reverse) process, we then have $\Delta s = \beta(\Delta E-\Delta F)$:
\begin{align}
\frac{P'_f(Q,\Delta E)}{P'_r(-Q,-\Delta E)} = e^{\beta(Q+\Delta E-\Delta F)}.
\label{eq:DFT:QE}
\end{align}
Alternatively, using the first law $\Delta E = W - Q$, we can also write \cite{cro98,cro00}
\begin{align}
\frac{P'_f(Q,W)}{P'_r(-Q,-W)} &= e^{\beta(W-\Delta F)}.\nn
\end{align}
A brief derivation is as follows \cite{noh12_prl}:
\begin{align}
P'_r(-Q) &= \int dW~P'_r(-Q,-W) \nn\\
&=  \int dW~P'_f(Q,W) e^{-\beta(W-\Delta F)} \nn\\
&= e^{\beta \Delta F} P'_f(Q) \int dW~P'_f(W|Q) e^{-\beta W} \nn\\
&= e^{\beta \Delta F} P'_f(Q) \av{e^{-\beta W}|Q}'_f.
\end{align}
Thus, we have
\begin{align}
 \frac{P'_f(Q)}{P'_r(-Q)} &= \frac{e^{-\beta \Delta F}}{\av{e^{-\beta W}|Q}'_f}.
\label{eq:DFT_QW}
\end{align}
On the other hand, if the system begins and ends in a NESS, then eq. \eqref{eq:DFT_traj} gives  \cite{sei05,sei08,cro99}
\begin{align}
\frac{P_f^{ss}(Q,\Delta s)}{P_r^{ss}(-Q,-\Delta s)} = e^{\beta Q+\Delta s} ~~\Ra~~\frac{P_f^{ss}(Q)}{P_r^{ss}(-Q)} &=\frac{e^{-\beta Q}}{\av{e^{-\Delta s}|Q}^{ss}_f} .
\label{eq:DFT_Qs}
\end{align}
The symbols $\av{\cdots}'$ and $\av{\cdots}^{ss}$ have been explained in the last section. The conditional averages appearing in the denominators on RHS have been calculated over all trajectories along which the same amount of heat has been released into the bath. To further clarify this point, we explicitly write the definitions appearing in eqs.  \eqref{eq:DFT_QW} and \eqref{eq:DFT_Qs} 
as follows. 
\emph{$\av{e^{-\beta W}|Q}'_f$ is the average of the quantity $e^{-\beta W}$, over all forward  trajectories  characterized by the fixed value $Q$ of dissipated heat, given that the initial points of these trajectories have been sampled from the equilibrium distribution}. 
On the other hand, \emph{$\av{e^{-\Delta s}|Q}^{ss}_f$ is the is the average of the quantity $e^{-\Delta s}$, over all forward  trajectories  characterized by the fixed value $Q$ of dissipated heat, given that the initial and final points on the trajectory follow steady state distributions corresponding to the instantaneous values of the protocols (given by $\lambda(0)$ and $\lambda(\tau)$, respectively).}

\subsection{MDFT for $\Delta s$ and $\Delta E$}

For a system in a NESS, or going from one steady state to another, the relation $\frac{P_f(Q,\Delta s)}{P_r(-Q,-\Delta s)} = e^{\beta Q+\Delta s}$ holds. A simple
cross-multiplication and integration over heat $Q$ then finally gives
%
\begin{comment}
  \int dQ ~P(Q,\Delta s) e^{-\beta Q} &= \int dQ~P(-Q,-\Delta s)e^{\Delta s} \nn\\
  \Ra P(\Delta s)\int dQ~P(Q|\Delta s)e^{-\beta Q} &= P(-\Delta s)e^{\Delta s}\int dQ~P(-Q|-\Delta s) \nn\\
	    &= P(-\Delta s)e^{\Delta s}
	\end{comment}
%
\begin{align}
	\Ra \frac{P_f^{ss}(\Delta s)}{P_r^{ss}(-\Delta s)} &= \frac{e^{\Delta s}}{\av{e^{-\beta Q}|\Delta s}_{ss}}.
	\label{eq:DFT_s}
\end{align}
\noindent As mentioned earlier, the above derivation applies only to processes that start and end in nonequilibrium steady states or in equilibrium states.

Similarly, for a system starting from an initial equilibrium state,  rewriting the Jarzynski equality as $\frac{P'_f(Q,\Delta E)}{P'_r(-Q,-\Delta E)} = e^{\beta (Q+\Delta E-\Delta F)}$, we can show that 
\begin{comment}
\frac{P_f(W)}{P_r(-W)} &= e^{\beta (W-\Delta F)} 
~~\Ra~~ \frac{P_f(Q,\Delta E)}{P_r(-Q,-\Delta E)} = e^{\beta (Q+\Delta E-\Delta F)} \nn\\
\end{comment}
%
\begin{align}
\Ra \frac{P'_f(\Delta E)}{P'_r(-\Delta E)} &= \frac{e^{\beta (\Delta E-\Delta F)}}{\av{e^{-\beta Q}|\Delta E}'_f}.
\label{eq:DFT_E}
\end{align}
We note that this theorem holds only when the system begins in an equilibrium state.

\section{Integral relations}

A collection of integral relations that can be obtained from the trajectory-level DFTs for work and total entropy:
\begin{align}
\mbox{Equilibrium distribution at initial time:}\hspace{1cm}\frac{P'_f[X]}{P'_r[\bar X]} &= e^{\beta (W[X]-\Delta F)}=e^{\beta (Q[X]+\Delta E(x_0,x_\tau)-\Delta F)}; \nn\\
 \mbox{Arbitrary distribution at initial time:}\hspace{1cm} \frac{P_f[X]}{P_r[\bar X]} &= e^{\Delta s_{tot}[X]} = e^{\beta Q[X]+\Delta s(x_0,x_\tau)}.
 % = e^{\beta (W-\Delta E)+\Delta s}.
\end{align}
Note that in the second equation, we do not need the restriction that the initial and final distributions must be stationary distributions. Thus,  the IFT corresponding to this trajectory-level DFT is a very general one:
%
%\begin{subequations}
\begin{align}
\av{e^{-\Delta s(x_0,x_\tau)}}_f &= \av{e^{-\beta Q[X]}}_r,
\end{align}
%\end{subequations}
for arbitrary initial states. 
In particular, if the initial state is an equilibrium one for both processes, then $\Delta s = \beta(\Delta E-\Delta F)$, and we get
\begin{align}
\av{e^{-\beta\Delta E(x_0,x_\tau)}}'_f &= e^{-\beta\Delta F}\av{e^{-\beta Q[X]}}'_r.
\end{align}
We note that although the MDFTs are difficult to test experimentally, the experimental verification of the MIFTs should be simpler.

\section{New MDFT for arbitrary initial states}

The conventional fluctuation theorems involving the nonequilibrium work $W$, are derived on the basis of the fact that the system is at equilibrium to begin with, after which a time-dependent protocol drives it away from the equilibrium state. The more general form of fluctuation theorem for the ratio of forward to reverse trajectories, involves the total entropy change $\Delta s_{tot}$ of the system. The dissipated work, $W-\Delta F$, coincides with $\Delta s_{tot}$ \emph{only} when the system starts from equilibrium. If not, then we need a free energy \emph{different} from the one mentioned above, to equate the dissipated work and total entropy change. We call this quantity the 
``nonequilibrium free energy'', defined by $F_{neq}(x,t)=E-Ts$. Then, using the definition of the equilibrium distribution $p^{eq}(x,t) = \frac{e^{-\beta E(x)}}{Z(t)}$, one can readily show that at any time instant $t$, we have \cite{lut12_arxiv,esp11_epl,lah09_pre}
\begin{align}
F_{neq}(x,t) &= E(x,t)-Ts(x,t) \nn\\
&= F(t)-T\ln p^{eq}(x,t) + T\ln p(x,t) \nn\\
&= F(t) + T\ln\frac{p(x,t)}{p^{eq}(x,t)}.
\end{align}
Here, $F(t)\equiv -T\ln Z(t)$ is the \emph{equilibrium} free energy corresponding to the value of the drive at time $t$. 
We denote the last term as
\begin{align}
D(x,t) = \ln\frac{p(x,t)}{p^{eq}(x,t)}.
\label{def:D}
\end{align}
Averaging $D(x,t)$ over the instantaneous distribution $p(x,t)$ gives the Kullback-Leibler divergence between the instantaneous distribution $p(x,t)$ and the equilbrium distribution $p^{eq}(x,t)$, which is defined as
\begin{align}
D_{KL}[p(x,t)||p^{eq}(x,t)] = \int dx ~p(x,t)\ln\frac{p(x,t)}{p^{eq}(x,t)}.
\label{def:DKL}
\end{align}

Using these definitions, and defining $\Delta F_{neq}(x_0,x_\tau) = F_{neq}(x_\tau,\tau)-F_{neq}(x_0,0)$, one can write the total entropy as \cite{lut12_arxiv,esp11_epl}
\begin{align}
\Delta s_{tot}[X] = \beta (W[X]-\Delta F_{neq}(x_0,x_\tau)) &= \beta (W[X]-\Delta F) - D(x_\tau,\tau)+ D(x_0,0) \nn\\
&= \beta (W[X]-\Delta F) - \Delta D(x_0,x_\tau),
\label{eq:SW_relation}
\end{align}
where we have defined
\begin{align}
\Delta D(x_0,x_\tau) \equiv D(x_\tau,\tau)- D(x_0,0).
\end{align}
All the fluctuation theorems derived above can be case entirely in terms of these relations, by using 
\begin{align}
\frac{P_f[X]}{P_r[\bar X]} = e^{\beta (W[X]-\Delta F) - \Delta D(x_0,x_\tau)}.
\end{align}
We immediately obtain the IFT
\begin{align}
\av{e^{-\beta (W[X]-\Delta F) + \Delta D(x_0,x_\tau)}}_f=1,
\end{align}
and application of Jensen's inequality gives
\begin{align}
\av{W[X]}\ge \Delta F+T\av{\Delta D(x_0,x_\tau)}_f.
\label{ineq:D}
\end{align}
From eqs. \eqref{def:D} and \eqref{def:DKL}, we find that $\av{\Delta D(x_0,x_\tau)}_f$  is simply the difference between the relative entropies at the beginning and at the end of the process \cite{lut12_arxiv,esp11_epl}
\begin{align}
\av{\Delta D(x_0,x_\tau)}_f = D_{KL}[p(x_\tau,\tau)||p^{eq}(x_\tau,\tau)] - D_{KL}[p(x_0,0)||p^{eq}(x_0,0)].
\end{align}
Note that \eqref{ineq:D} is a different inequality as compared to the Jarzynski equality, since in the latter case the second term on the RHS was absent. 
%In case $\av{\Delta D}$ becomes negative, it is then possible to extract more work (compared to the free energy change) from the system. 
Eq. \eqref{eq:SW_relation}, when averaged, gives
\begin{align}
\av{W}=\Delta F + T[\av{\Delta s_{tot}}+\av{\Delta D}]. 
\label{av_SW}
\end{align}
Comparing \eqref{ineq:D} and \eqref{av_SW}, we obtain as a corollary the inequality, $\av{\Delta s_{tot}}\ge 0$, which is essentially the second law for mesoscopic systems \cite{sei12}.
We further observe that work higher than $\Delta F$ can be extracted from the system,  when the condition $\av{\Delta D}<-\av{\Delta s_{tot}}$ holds.

We further note that the following integral relation can be derived:
\begin{align}
\av{e^{-\beta (W[X]-\Delta F)}}_f &= \av{e^{-\Delta D(x_0,x_\tau)}}_r=1.
\end{align}
The last equality follows from the Jarzynski equality $\av{e^{-\beta (W[X]-\Delta F)}}_f=1$. Further, if the system begins and ends in steady states, then $\Delta D$ changes sign in the reverse process, and we get
\begin{align}
\av{e^{\Delta D_r(x_\tau,x_0)}}^{ss}_r =1.
\end{align}
Since the IFTs must be valid for both the forward and the reverse processes, we can write
\begin{align}
\av{e^{\Delta D_f(x_0,x_\tau)}}^{ss}_f=1.
\end{align}

Here, we have used the fact that the signs of $W$ and $\Delta F$ change for the reverse trajectory. 
All the relations would be very general and on equal footing as those obtained from \eqref{eq:DFT_traj}.
As in the case of system entropy, $\Delta D(x_0,x_\tau)$ does not in general change sign on time-reversal, because the initial and final instantaneous distributions do not interchange their forms in the reverse process. However, they do so when the end points of the trajectory follow steady state (or equilibrium) distributions. In this case, the DFT can be written as (compare with eq. \eqref{eq:DFT_Qs})
\begin{align}
\frac{P^{ss}_f(W,\Delta D)}{P^{ss}_r(-W,-\Delta D)} = e^{\beta (W-\Delta F) - \Delta D}.
\end{align}
Here, $P^{ss}_f(W,\Delta D)$ is the probability of $W[X]$ taking a specific value $W$, and $\Delta D(x_0,x_\tau)$ taking a specific value $\Delta D$ (see eq. \eqref{eq:basic}).
The MDFT for $\Delta D$ can be obtained as
\begin{align}
\frac{P^{ss}_f(\Delta D)}{P^{ss}_r(-\Delta D)} = \frac{e^{-\Delta D-\beta\Delta F}}{\av{e^{-\beta W}|\Delta D}^{ss}_f}.
\end{align}
This equation may be compared to \eqref{eq:DFT_s}. An alternative form may be derived by starting with (using the first law: $W=Q+\Delta E$)
\begin{align}
\frac{P^{ss}_f(Q,\Delta E,\Delta D)}{P^{ss}_r(-Q,-\Delta E,-\Delta D)}& = e^{\beta (Q+\Delta E-\Delta F) - \Delta D}.
%\Ra P^{ss}_f(\Delta E) P^{ss}_f(Q,\Delta D|\Delta E)& = e^{\beta(\Delta E-\Delta %F)} P^{ss}_r(-\Delta E) P^{ss}_r(-Q,-\Delta D|-\Delta E)e^{\beta (Q-\Delta D)}.
\end{align}
%
%Integration of both sides over $Q$ and $\Delta D$ yields
%
which leads to the MDFT
\begin{align}
\frac{P^{ss}_f(\Delta E)}{P^{ss}_r(-\Delta E)} &= \frac{e^{\beta(\Delta E-\Delta F)}}{\av{e^{-\beta (Q-\Delta D)}|\Delta E}^{ss}_f}.
\label{eq:DFT_E_ss}
\end{align}
Comparing this equation with \eqref{eq:DFT_E}, we find two differences:
\begin{enumerate}
\item The LHS contains ratio of steady state probabilities in eq. \eqref{eq:DFT_E_ss}. That is why the subscripts $f$ and $r$ are no longer present.

\item The denominator in the RHS of \eqref{eq:DFT_E_ss} contains steady state averages, and the argument within the average is different.
\end{enumerate}

Table \ref{table} summarizes the relations obtained up to now. Similar results can be obtained for the quantum system as well, under the assumption of weak coupling between the system and the heat bath. The steps leading to the DFT for total entropy change have been outlined in the appendix. As an example, we have shown how the MDFT and MIFT for system entropy change follow from this relation. Other relations can be obtained using similar mathematical treatment.

%\vspace{0.5cm}

\begin{center}
\begin{table}[!h]
\caption{\bf Summary of the results}
\begin{tabular}{|p{3cm}|l|}
\hline
&\\
Transient MDFTs & $\dfrac{P'_f(Q)}{P'_r(-Q)} = \dfrac{e^{-\beta \Delta F}}{\av{e^{-\beta W}|Q}'_f}$ \\
&\\
 & $\dfrac{P'_f(\Delta E)}{P'_r(-\Delta E)} = \dfrac{e^{\beta (\Delta E-\Delta F)}}{\av{e^{-\beta Q}|\Delta E}'_f}$\\
 &\\
 \hline
 &\\
  MDFTs for initial and final stationary states& $\dfrac{P^{ss}_f(Q)}{P^{ss}_r(-Q)}= \dfrac{e^{-\beta Q}}{\av{e^{-\Delta s}|Q}^{ss}_f}$ \\
 &\\
 & $\dfrac{P^{ss}_f(\Delta s)}{P^{ss}_r(-\Delta s)} = \dfrac{e^{\Delta s}}{\av{e^{-\beta Q}|\Delta s}^{ss}_f}$ \\
 &\\
 & $\dfrac{P^{ss}_f(\Delta D)}{P^{ss}_r(-\Delta D)} = \dfrac{e^{\Delta D-\beta\Delta F}}{\av{e^{-\beta W}|\Delta D}^{ss}_f}$\\
 &\\
 & $\dfrac{P^{ss}_f(\Delta E)}{P^{ss}_r(-\Delta E)} = \dfrac{e^{\beta(\Delta E-\Delta F)}}{\av{e^{-\beta (Q+\Delta D)}|\Delta E}^{ss}_f}$\\
 &\\
 \hline
 &\\
MIFTs & $\av{e^{-\Delta s}}_f = \av{e^{-\beta Q}}_r$\\
 &\\
&$ \av{e^{-\beta\Delta E}}'_f = e^{-\beta\Delta F}\av{e^{-\beta Q}}'_r$\\
&\\
& $ \av{e^{-\beta (W-\Delta F)+\Delta D}}_f=1$\\
&\\
&$\av{e^{\Delta D}}^{ss}_f=1$.\\
\hline
\end{tabular}
\label{table}
\end{table}
\end{center}

\section{Presence of information}

The thermodynamic quantities, like work or total entropy change, that have \emph{exact} detailed fluctuation theorems, have integral fluctuation theorems of the form:
\begin{align}
\av{e^{-\beta(W[X]-\Delta F)}}'_f &= 1. \nn\\
\av{e^{-\Delta s_{tot}[X]}}_{f} &=1.
\end{align}
Such equations have been generalized to case of feedback-controlled systems \cite{sag10_prl,lah12_jpa,sag12_pre}. Here, application of feedback to the system is defined in the following sense. We first measure the state of the system at time $t=0$, where the system is actually in the state $x_0$. However, due to inaccuracy of measurement, we obtain the outcome $m_0$ with the error probability $p(m_0|x_0)$. We now apply the protocol $\lambda_{m_0}(t)$ from time $t=0$ to $t=t_1$, when we make another measurement of the system state. We obtain the outcome $m_1$ with probability $p(m_1|x_1)$, and apply the protocol $\lambda_{m_1}(t)$, and so on. let there be $N$ such measurements in total. There will be many protocols generated in the process, and we can choose any one of them and call it as the protocol for the ``forward process''. The ``reverse process'' can then be defined as the one where this particular protocol is blindly time-reversed. Let the sequence of measurements $\{m_0,m_1,\cdots,m_N\}$ be denoted by $M$. Then the generalized IFTs become
\begin{align}
\av{e^{-\beta(W[X,M]-\Delta F(m_0,m_N))-I[X,M]}}'_f &= 1; \nn\\
\av{e^{-\Delta s_{tot}[X,M]-I[X,M]}}_{f} &=1,
\label{eq:EFT}
\end{align}
where the mutual information $I$ is defined as 
\begin{align}
I[X,M] \equiv \frac{p(m_0|x_0)p(m_1|x_1)\cdots p(m_N|x_N)}{p(m_0,m_1,\cdots,m_N)}.
\label{def:I}
\end{align}
The ensemble averages have been carried out over all phase-space trajectories $X$ and all measurement trajectories $M$.
Application of Jensen's inequality then gives the modified second laws
\begin{align}
\av{W[X,M]-\Delta F(m_0,m_N)}'_f \ge -\av{I[X,M]}; \nn\\
\av{\Delta s_{tot}[X,M]}_{f}\ge -\av{I[X,M]}.
\end{align}
This means that in principle, extraction of work (exceeding $\Delta F$) is possible in presence of information, if the feedback algorithm is efficient enough.

However, when the DFTs are not exact, we do not have such modified second laws, where in principle work can be extracted from the system. For instance, let us consider the MDFT for heat:
\begin{align}
\frac{P^{ss}_f(Q)}{P^{ss}_r(-Q)} = \frac{e^{\beta Q}}{\av{e^{-\Delta s}|Q}^{ss}_f}.
\end{align}
In presence of information, this MDFT gets modified to 
\begin{align}
\frac{P^{ss}_f(Q,I)}{P^{ss}_r(-Q,I)} = \frac{e^{\beta Q+I}}{\av{e^{-\Delta s}|Q}^{ss}_f}.
\end{align}
Here, $P(Q,I)$ is the joint distribution of the  heat dissipated and the mutual information gained during the process \cite{sag10,hor10}.
The corresponding IFT will be 
\begin{align}
\av{e^{-\beta Q -I}}^{ss}_f = \av{e^{-\Delta s}}^{ss}_r.
\end{align}
This can be readily read off from the trajectory-level DFT \cite{lah12_jpa}
\[
\frac{P_f[X,M]}{P_r[\bar X;M]}=e^{{\beta Q+\Delta s+I}},
\]
keeping in mind that $\Delta s$ changes sign in the reverse process only when the process begins and ends in steady states.
Jensen's inequality gives
\begin{align}
\av{e^{-\Delta s}}^{ss}_f \ge e^{-\av{\beta Q+I}^{ss}_f} ~~\Ra~~ \av{Q}^{ss}_f \ge -k_BT[\ln \av{e^{-\Delta s}}^{ss}_f + \av{I}^{ss}_f],
\label{eq:no_law2}
\end{align}
which says nothing about the positivity of mean heat. Thus, second-law-like inequalities cannot be formulated for the thermodynamic variables that follow MDFT instead of an exact DFT.

\subsection{Comment on the extended fluctuation theorems under information gain}

In general, the relations \eqref{eq:EFT} are not unique. The  correction term, given by eq. \eqref{def:I}, is valid if the reverse process is generated by simply time-reversing one of the forward trajectories \cite{lah12_jpa,lah13_phyA}.
 In fact, if the reverse process is not generated by a simple time-reversal of the forward protocol, but is generated by other methods as described in \cite{kun12_pre,lah13_phyA}, we can have other expressions for this correction term. One such method is to apply feedback along the reverse process as well.
 Suppose we have a sequence of measurement given by $\{m_0,m_1,\cdots,m_{N}\}$ at times $\{t_0,t_1,\cdots,t_{N}\}$, which defines the forward process (forward protocol). To respect causality, the exact reverse protocol will correspond to the set of measurements $\{m_{N},m_{N-1},\cdots,m_0\}$ at the \emph{shifted} time instants $\{t_{N+1},t_N,\cdots,t_1\}$. In this case, the mutual information $I$ apearing in \eqref{def:I} will be replaced by 
 \begin{align}
 \phi &= \frac{p(m_0|x_0)p(m_1|x_1)\cdots p(m_N|x_N)}{p(m_0|x_1)p(m_1|x_2)\cdots p(m_N|x_{N+1})}.
 \end{align}
 Likewise, using a combination of both the methods described above for generating the reverse protocol, various different correction terms appear \cite{lah13_phyA}.
  Nevertheless, the conclusion below eq. \eqref{eq:no_law2} remains unaltered.

\section{Conclusions}

In conclusion, in this paper we have used the fluctuation theorem \eqref{eq:DFT_traj0}, to generate relations that resemble the detailed fluctuation theorems, saving the fact that an extra term appears in the relation. We call them the modified detailed fluctuation theorems or MDFTs. Similarly, we also obtain a few modified theorems in their integral forms. These relations contain the heat $Q$, internal energy change $\Delta E$, system entropy change $\Delta s$, relative entropy change $\Delta D$, etc. They are not very common in literature, since they do not lead to any useful inequality like the second law. Nevertheless, the derivations show that such relations can be obtained for many different thermodynamic quantities. Such relations can also be derived for the so-called housekeeping and excess heats \cite{oon98,hat01_prl,esp10_prl,lah14_arxiv}, as well as for exchanged heat with a system connected to two reservoirs \cite{lah14_arxiv}, but the algebra is similar and has not been reproduced here. Unlike the MDFTs, experimental verification of MIFTs should be simpler.

\section{Acknowledgement}

One of us (AMJ) thanks DST, India for financial support.

\appendix
\section{Derivations for the quantum case}

We now briefly consider the extension of the results to a quantum system that is interacting with a heat bath. Let the Hamiltonians for the system, the bath and the interction force be denoted by $H_S(t)$, $H_B$ and $H_{SB}$, respectively. We assumed that only the system Hamiltonian depends explicitly on time, because of the time-dependence of the external perturbation. We will express the path probability in terms of state vectors (see eq. \eqref{pf_quantum}), rather than using the equivalent density matrix approach \cite{han09_jsm}.

Let us consider a quantum system that is weakly correlated to a heat bath that is held at temperature $T$. The total Hamiltonian is given by
\begin{align}
H(t) = H_S(t)+H_B+H_{SB}.
\end{align}
The combined supersystem is initially (time $t=0$) at thermal equilibrium:
\begin{align}
\rho(0) = \frac{e^{-\beta H(t)}}{Z_0}.
\end{align}
%

%%%%%%%%%%%%%%%%%%%%%%%%%%%%%%%%%%
\begin{comment}
We perform the first measurement on the system at $t=0$. Then the state immediately after measurement 
\[
\bar{\rho}(0) = \sum_{i,\alpha}P_{i\alpha}\rho(0)P_{i\alpha},
\]
where $i$ and $\alpha$ refer to system and bath states, respectively. $P_{i\alpha}$ is the projection operator: $P_{i\alpha} = |i,\alpha\ra\la i,\alpha|$.

If $H_{SB}\ll H_S(t),H_B$, then we have,
%
\begin{align}
\bar{\rho}(0) = \frac{e^{-\beta H_S(0)}}{Z_S(0)}\cdot \frac{e^{-\beta H_B}}{Z_B}.
\end{align}
%
\end{comment}
%%%%%%%%%%%%%%%%%%%%%%%%%%%%%%%%%%

If $H_{SB}\ll H_S(t),H_B$, then the initial probability (at $t=0+$) of the states $i_0$ and $\alpha_0$ after simultaneous projective measurements are performed on the states of the system and bath \cite{han09_jsm}, is given by
\begin{align}
p_{i_0\alpha_0} &= \Tr{S,B}[\Pi_{i_0\alpha_0}~\rho(0)] \simeq \frac{e^{-\beta E_i}}{Z_S(0)}~ \frac{e^{-\beta E_\alpha}}{Z_B}.
\end{align}
Here, $\Pi_{i_0\alpha_0}\equiv |i_0,\alpha_0\ra\la i_0,\alpha_0|$ is the projection operator.

 If the system+bath goes from states $|i_0,\alpha_0\ra$ to the states $|i_\tau,\alpha_\tau\ra$, then we define the path probability as
 \begin{align}
 P_f(i_0\alpha_0\to i_\tau\alpha_\tau) = \pr{\tau}{0}~p_{i_0\alpha_0}, 
 \label{pf_quantum}
 \end{align}
 where 
 \begin{align}
 \pr{\tau}{0} = |\la i_{\tau},\alpha_{\tau}|U(\tau,0)|i_0,\alpha_0\ra|^2,
 \end{align}
 $U(\tau,0)$ being the unitary evolution operator between the time instants $0$ and $\tau$.
 Similarly,
 \begin{align}
 P_r(i_0\alpha_0\leftarrow i_\tau\alpha_\tau) =  \pr{0}{\tau}~p_{i_\tau\alpha_\tau}.
 \label{pr_quantum}
 \end{align}
Since $\pr{\tau}{0}=\pr{0}{\tau}$ \cite{lah12_jpa,lah12_pramana}, we immediately get
\begin{align}
\frac{P_f(i_0\alpha_0\to i_\tau\alpha_\tau)}{P_r(i_0\alpha_0\leftarrow i_\tau\alpha_\tau)}= e^{\beta(E_{i_\tau}-E_{i_0})+\beta(E_{\alpha_\tau}-E_{\alpha_0}-\beta(F_S(\tau)-F_S(0))}.
\end{align}
Defining $\Delta E \equiv E_{i_\tau}-E_{i_0}$, $Q \equiv E_{\alpha_\tau}-E_{\alpha_0}$ and $\Delta F \equiv F_S(\tau)-F_S(0)$, we get
\begin{align}
\frac{p_f(i_0\alpha_0\to i_\tau\alpha_\tau)}{p_r(i_0\alpha_0\leftarrow i_\tau\alpha_\tau)} = e^{\beta(\Delta E+Q-\Delta F)}.
\end{align}
Then we have,
\begin{align}
P_f(\Delta E,Q) &= \sum_{i_0,\alpha_0, i_\tau,\alpha_\tau}P_f(i_0\alpha_0\to i_\tau\alpha_\tau)~\delta(E_{i_\tau}-E_{i_0}-\Delta E)~\delta(E_{\alpha_\tau}-E_{\alpha_0}-Q) \nn\\
&= e^{\beta(\Delta E+Q-\Delta F)}\sum_{i_0,\alpha_0, i_\tau,\alpha_\tau} P_r(i_0\alpha_0\leftarrow i_\tau\alpha_\tau)~\delta(E_{i_\tau}-E_{i_0}-\Delta E)~\delta(E_{\alpha_\tau}-E_{\alpha_0}-Q) \nn\\
&= e^{\beta(\Delta E+Q-\Delta F)}\sum_{i_0,\alpha_0, i\tau,\alpha_\tau} P_r(i_0\alpha_0\leftarrow i_\tau\alpha_\tau)~\delta(E_{i_0}-E_{i_\tau}+\Delta E)~\delta(E_{\alpha_0}-E_{\alpha_\tau}+Q) \nn\\
&=P_r(-\Delta E,-Q)~e^{\beta(\Delta E+Q-\Delta F)}.
\label{eq:DFT_QE}
\end{align}
Then we can write,
\begin{align}
P_r(-Q) &= \int dW P_r(-Q,-W) = \int dW~ P_f(Q,W) ~e^{-\beta (W- \Delta F)}\nn\\
&=e^{\beta\Delta F} P_f(Q) \int dW~ p_f(W|Q)~e^{-\beta W} \nn\\
\Ra &\boxed{ 
\frac{P_f(Q)}{P_r(-Q)} = \frac{e^{-\beta\Delta F}}{\la e^{-\beta W}|Q\ra}.
}
\end{align}
Of course, if in equations \eqref{pf_quantum} and \eqref{pr_quantum}, we had considered the initial probability distribution of the system for the forward process to be arbitrary rather than the equilibrium one, and that of the reverse process to be the final distribution of the foward process, while the bath is always at equilibrium, then we would have obtained the fluctuation relation
\begin{align}
\frac{p_f(i_0\alpha_0\to i_\tau\alpha_\tau)}{p_r(i_0\alpha_0\leftarrow i_\tau\alpha_\tau)} = e^{\Delta s_{tot}}=e^{\beta Q+\Delta s},
\label{pathratio_quantum}
\end{align}
This is the quantum analogue of the classical relation given by eq. \eqref{eq:DFT_traj0} \cite{lah13_pramana}.

As is obvious, all the relations derived for the classical case can be similarly generalized to the quantum case, once we convert the trajectory ratio to the ratio of probability distributions for the thermodynamic variables. For example, the MDFT for  $\Delta s$ can be derived by converting eq. \eqref{pathratio_quantum} into the ratio of $P_f(Q,\Delta s)$ and $P_r(-Q,-\Delta s)$
\begin{align}
\frac{P_f(Q,\Delta s)}{P_r(-Q,-\Delta s)} &= e^{\beta Q+\Delta s} \nn\\
\Ra \frac{P_f(\Delta s)}{P_r(-\Delta s)}  &=  \frac{e^{\Delta s}}{\av{e^{-\beta Q}|\Delta s}_{ss}}.
\end{align}
Here we  use the same definition for the change in system entropy as in the classical case, namely the logarithm of the ratio of initial to the final distribution: $\Delta s\equiv \ln[p_{i_0\alpha_0}/p_{i_\tau\alpha_\tau}]$. 
Using \eqref{pathratio_quantum}, we can easily derive the MIFT
\begin{align}
\av{e^{-\beta Q}}_f^{ss} = \av{e^{-\Delta s}}^{ss}_r,
\end{align}
where we have once again taken note of the fact that $\Delta s$ switches sign in the reverse process, only when the either process begins and ends in stationary states.
Similar mathematics can be used to derive all the other relations listed in table \ref{table}.
All the relations remain valid even if intermediate measurements of arbitrary observables are performed on the system \cite{han10,lah12_pramana,lah13_pramana}. Finally, we note that the results in the quantum case may also be handled by using the concept of heat and work steps, and given in \cite{qua08_arxiv,lah13_pramana}.

\bibliographystyle{apsrev4-1}
\bibliography{ref}

\end{document}